\documentclass{PoS}

\usepackage{amsfonts} 
\usepackage{amssymb} 
\usepackage{amsmath} 
\usepackage{graphicx} 
\usepackage{subcaption}
\usepackage{array} 
\usepackage{bm} 
\usepackage{longtable} 
\usepackage[dvipsnames]{xcolor} 
\usepackage{comment}
\usepackage{verbatim}
\usepackage{multirow}

\providecommand{\wbar}[1]{\overline#1}
\providecommand{\abs}[1]{\lvert#1\rvert}

\providecommand{\matrixe}[3]{\langle#1\lvert#2\rvert#3\rangle}
\providecommand{\expv}[1]{\langle#1\rangle}

\providecommand{\fm}{\mathrm{fm}}

\newcommand{\Dslash}{\ensuremath{{D\kern -0.65em /}}}

\providecommand{\MeV}{\mathrm{MeV}}

\definecolor{indigo}{HTML}{4B0082}
\definecolor{darkorange}{HTML}{FF8C00}
\definecolor{darkgreen}{HTML}{006400}
\definecolor{myred}{HTML}{ED3B3B}
\definecolor{mygreen}{HTML}{2E8B57}
\definecolor{myblue}{HTML}{2E74FF}
\definecolor{mydarkblue}{HTML}{1E0966}
\definecolor{mybgc}{HTML}{FFFACD}
\definecolor{fgray}{cmyk}{0.23,0.11,0.00,0.57}
\definecolor{fyellow}{cmyk}{0.00,0.35,1,00,0.00}
\definecolor{fgreen}{cmyk}{0.61,0.00,0.66,0.39}

\newcommand{\Vcb}{|V_{cb}|}

\newcommand{\BtoDst}{\bar{B} \to D^\ast \ell \bar{\nu}}
\newcommand{\BtoD}{\bar{B} \to D \ell \bar{\nu}}

\graphicspath{{./figs/}}

\allowdisplaybreaks

\title{Semileptonic $B \to D^{(\ast)} \ell\nu$ Decay Form
  Factors \\ using the Oktay-Kronfeld Action}

\ShortTitle{$B \to D^{(\ast)} \ell\nu$ Decay Form Factors using the
  OK Action}

\author{{Yong-Chull Jang}\\
  Physics Department, Brookhaven National Laboratory,
  Upton, NY 11973, USA\\
  E-mail: \email{ypj@bnl.gov}}

\author{{Sungwoo Park, Tanmoy Bhattacharya, Rajan Gupta}\\
  Theoretical Division T-2, Los Alamos National Laboratory,
  Los Alamos, NM 87545, USA\\
  E-mail: \email{sungwoo@lanl.gov},
  \, \email{tanmoy@lanl.gov},
  \, \email{rg@lanl.gov}}

\author{{Benjamin J. Choi, \speaker{Seungyeob Jwa},
    Sunkyu Lee, Weonjong Lee}\\
  Lattice Gauge Theory Research Center, CTP, and FPRD,
  Department of Physics and Astronomy,
  Seoul National University, Seoul 08826, South Korea\\
  E-mail: \email{wlee@snu.ac.kr}}

\author{{Jaehoon Leem}\\
  School of Physics, Korea Institute for Advanced Study,
  Seoul 02455, South Korea\\
  E-mail: \email{leemjaehoon@kias.re.kr}}

\author{LANL/SWME Collaboration}

\abstract{
We report recent progress in calculating semileptonic form factors for
the $\BtoDst$ and $\BtoD$ decays using the Oktay-Kronfeld (OK)
action for bottom and charm quarks.
We use the second order in heavy quark effective power counting
$\mathcal{O}(\lambda^2)$ improved currents in this work.
The HISQ action is used for the light spectator quarks.
We analyzed four $2+1+1$-flavor MILC HISQ ensembles with $a\approx
0.09\,\fm$, $0.12\,\fm$ and $M_\pi \approx 220\,\MeV$, $310\,\MeV$:
$a09m220$, $a09m310$, $a12m220$, $a12m310$.
Preliminary results for $B\to D^\ast\ell\nu$ decays form factor
$h_{A_1}(w)$ at zero recoil ($w=1$) are reported.
Preliminary results for $B\to D\,\ell\nu$ decays form factors $h_\pm(w)$
over a kinematic range $1<w<1.3$ are reported as well.
}

\FullConference{37th International Symposium on Lattice Field
  Theory - Lattice2019\\
  16-22 June 2019\\
  Wuhan, China}

\begin{document}

\section{Introduction}
\label{sec:intro}
Semileptonic decays $B \to D^{(\ast)}\ell\nu$ are interesting
processes, because these are probes of the Cabibbo-Kobayashi-Maskawa
(CKM) matrix element $\Vcb$ involving heavy, the charm and bottom,
quarks \cite{ Amhis:2016xyh}.
In the Standard Model (SM), the CKM matrix is unitary.
Tests of the unitarity, e.g., global unitarity triangle analysis,
using inputs from experiments and lattice calculations has become
tight as the inputs are determined precisely.
However, higher precision is required for a stringent test of SM in
$\varepsilon_K$ \cite{ Kim:2019vic, Bailey:2018feb, Bailey:2015tba},
which makes the importance of $\Vcb$ escalate even more.
Currently, $\Vcb$ shows a $3\sigma \sim 4\sigma$ difference between
inclusive and exclusive determination using the CLN parameterization.
Meanwhile the BGL parameterization gave the exclusive $\Vcb$ that is
consistent with the inclusive determination, recent analysis with
Belle untagged data for $B\to D^\ast\ell\nu$ decays shifts the BGL
result on top of the previous CLN result~\cite{ Kim:2019vic}.
Ratios, $R(D)$ and $R(D^\ast)$, of branching fractions of the
semileptonic decays $B \to D^{(\ast)}\tau\nu$ to $B \to
D^{(\ast)}\ell\nu,\, (\ell=e,\mu)$ are also interesting in the test of
the lepton flavor universality in the SM \cite{ Amhis:2016xyh}.
Thus, our focus is on a determination of form factors for $B \to
D^{(\ast)}\ell\nu$ decays in a sub-percent precision using an improved
heavy quark discretization: the Oktay-Kronfeld (OK) action \cite{
  Oktay:2008ex}.

The OK action \cite{Oktay:2008ex} further improves the Fermilab
action~\cite{ElKhadra:1996mp} by including dimension 6 and 7 bilinear
terms necessary for tree-level matching to QCD through order
$\mathcal{O}(\lambda^3)$ ($\lambda \approx \bm{p}a \approx
\Lambda/(2m_Q)$) in heavy quark effective theory (HQET) power
counting.
The improvement coefficients are perturbatively calculated by matching
on-shell amplitudes at tree-level between continuum QCD and lattice
QCD.
Heavy quark discretization error enters in
$\mathcal{O}(\Lambda/m_{b,c})^n$ so that the errors can be controlled
with large lattice spacings $am_0 > 1$.
In contrast, the Symanzik improved action has $\mathcal{O}(am_0)^n$
discretization error, and thus, $am_b \sim 1$ requires $a \sim
0.045\,\fm$.
At present, MILC HISQ ensemble is available down to $a \sim
0.03\,\fm$~\cite{Bazavov:2017lyh}.

Here, we present preliminary analysis on form factors $h_{A_1}(w)$ at
zero-recoil $w=1$ and $h_{\pm}(w)$ over a kinematic range $1<w<1.3$.
The differential decay rate for $B\to D^{(\ast)}\ell\nu$ is parameterized by a conventional form factor $\mathcal{F}(w)$ ($\mathcal{G}(w)$):
\begin{align}
\frac{d\Gamma}{dw}(B\to D^\ast\ell\nu)
& = \frac{G_F^2 M_{D^\ast}^3}{48\pi^3} (M_B - M_{D^\ast})^2 (w^2-1)^{1/2}
\chi(w) \abs{\eta_\text{EW}}^2 {\abs{V_{cb}}^2} {\abs{\mathcal{F}(w)}^2} \,,
\label{eq:B2Dstar-rate}
\\
\frac{d\Gamma}{dw}(B\to D\;\ell\nu)
&= \frac{G_F^2 M_{D}^3}{48\pi^3} (M_B + M_D )^2 (w^2-1)^{3/2} 
\abs{\eta_\text{EW}}^2 {\abs{V_{cb}}^2} {\abs{\mathcal{G}(w)}^2} \,.
\label{eq:B2D-rate}
\end{align}
The $\mathcal{F}(w)$ is factorized by the $h_{A_1}(w)$ leaving the
helicity amplitudes that are normalized at zero-recoil.  We anticipate
that the improved action and current could address the issue in the $B
\to D^\ast\ell\nu$ form factor parametrizations.  The $\mathcal{G}(w)$
is a linear combination of the two form factors $h_{\pm}(w)$.

The improvement by the OK action was explicitly demonstrated by
calculating spectrum of heavy-light mesons and quarkonium \cite{
  Bailey:2017nzm}.
The corrections in $\alpha_s$ are partially taken into account by the
tadpole improvement for both the action and the current \cite{
  Bailey:2020uon}.
Improved current operator $J^{fg}$ for a flavor change $f \to g$ can
be written in terms of an improved (rotated) field $\Psi = \mathcal{R}
\psi$: $J^{fg} = \wbar{\Psi}^g \Gamma \Psi^f$.
Tree-level matching up to $\mathcal{O}(\lambda^3)$ is given in \cite{
  Bailey:2020uon}.

Three-point correlation functions $C(t,\tau)$
\begin{align}
  C^{X(f)\to Y(g)}(t,\tau)
  =& \expv{\mathcal{O}_{Y}^\dagger(0) J^{fg}(t) \mathcal{O}_X(\tau)}
  \\=& \mathcal{A}_{Y} \mathcal{A}_X \frac{\matrixe{Y}{J^{fg}}{X}}{\sqrt{2M_Y}\sqrt{2M_X}} e^{-M_{Y}t} e^{-M_X(\tau-t)} + \cdots
\end{align}
is calculated for multiple, four to six, source and sink separations
$\tau$.
Then, the correlators are analyzed with multistate fits to extract the
matrix elements~\cite{Bhattacharya:2018ibo}, which are decomposed into
the form factors.
See a companion proceeding \cite{ Bhattacharya:2020ahq} for the rest
of details about tuned quark masses, smearing, the $2+1+1$-flavor MILC
HISQ ensembles used in this work, definitions of the OK action and the
improved heavy quark field.
%

\section{Meson Spectrum}
\label{sec:spec}
Meson masses can be obtained from the kinetic mass $M_2$, which is
extracted from fit to the dispersion relation
\begin{equation}
  E = M_1 + \frac{\bm{p}^2}{2M_2}
  - \frac{(\bm{p}^2)^2}{8M_4^3} - \frac{a^3w_4 }{6}\sum_i p_i^4 + \cdots\,.
\end{equation}

%
%
%
\begin{figure}[t!]
  \begin{subfigure}{0.485\linewidth}
    \includegraphics[width=\linewidth]{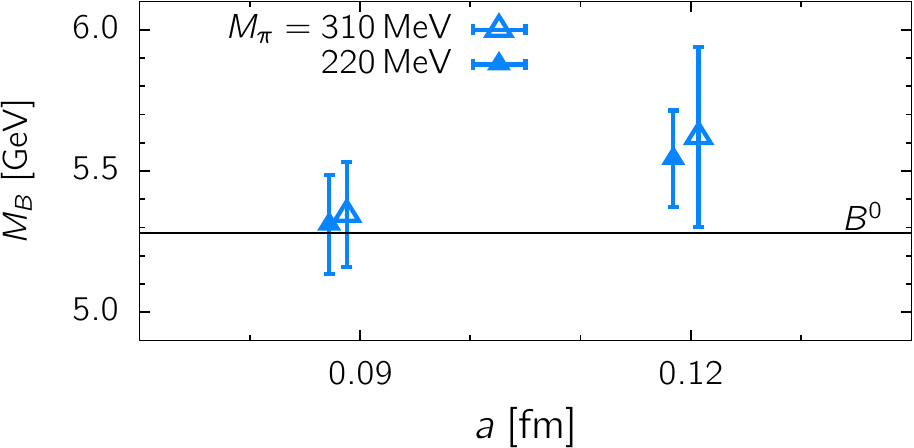}
    \caption{$B$ meson mass}
    \label{fig:B-mass}
  \end{subfigure}
  \hfill
  \begin{subfigure}{0.485\linewidth}
    \includegraphics[width=\textwidth]{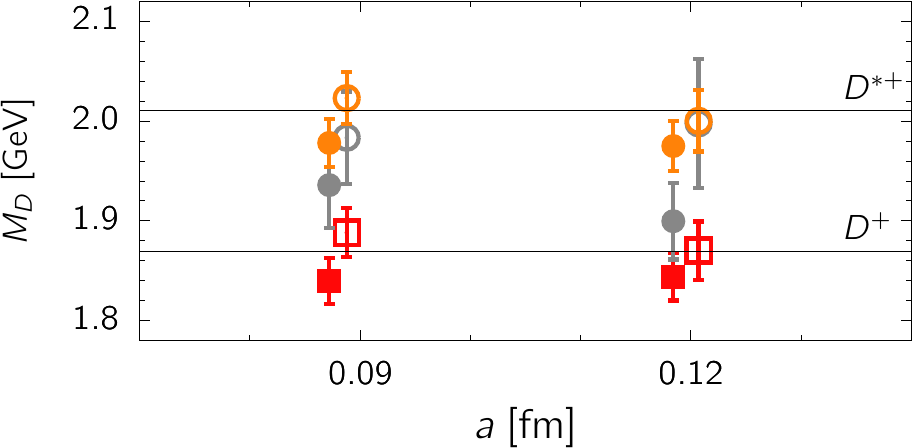}
    \caption{$D^+$ and $D^\ast$ meson masses}
    \label{fig:D-Dst-mass}
  \end{subfigure}
  \caption{Masses of $B^{0}$, $D^{\ast}$ and $D^{+}$.  The horizontal
    lines correspond to the physical masses from PDG. The orange
    (gray) circle represents the (alternative) kinetic meson mass
    $M_2(D^\ast)$. The filled (open) symbol corresponds to the $M_\pi
    = 220\,\MeV$ ($310\,\MeV$).}
  \label{fig:meson-spec}
\end{figure}

In Fig.~\ref{fig:meson-spec}, the masses of $B$, $D^{+}$, $D^{*}$
mesons from four ensembles are compared to the physical masses from
PDG.
Alternatively, $M_2(D^\ast) = M_2(D) + M_1(D^\ast) - M_1(D)$ (the
orange circles) gives another $D^\ast$ mass, because the mass
splitting between the rest masses is a physical quantity.
This alternative kinetic mass of $D^\ast$ is better consistent with
the physical mass for all four ensembles.
The rest mass is solely determined by the zero momentum meson
correlator, and thus results in a smaller error than the $M_2(D^\ast)$
determined from the fitting.  Note that the vector meson correlator is
noisier than the pseudoscalar meson correlator for a given momentum
$\bm{p}$.
%

\section{$B\to D^\ast\ell\nu$ form factor at zero recoil: $h_{A_1}(1)$}
\label{sec:h_A1-1}
The form factor $h_{A_1}(w=1)$ is obtained by taking the following
double ratio:
\begin{align}
  \rho_{A_1}^2 & \frac{\matrixe{B}{A^{bc}}{D^\ast}\matrixe{D^\ast}{A^{cb}}{B}}
      {\matrixe{B}{V^{bb}}{B}\matrixe{D^\ast}{V^{cc}}{D^\ast}}
      = \abs{h_{A_1}(1)}^2 \,,\quad
      \rho_{A_1}^2 =\frac{Z_A^{bc}Z_A^{cb}}{Z_V^{bb}Z_V^{cc}} \,.
\end{align}
\begin{figure}[tbp]
  \begin{center}
    \includegraphics[width=0.6\linewidth]{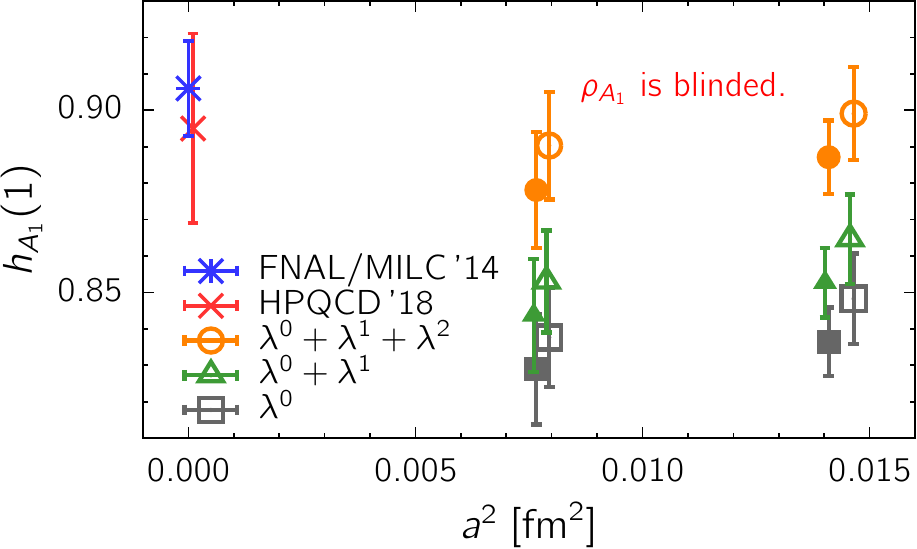}
  \end{center}
  \caption{$h_{A_1}(w)$ at zero recoil ($w=1$) from four
    ensembles. Current improvements up to order
    $\mathcal{O}(\lambda^n)$, $n=0,1,2$, are compared for each
    ensemble. The $n=0$ is with the unimproved current.  The filled
    (open) symbol corresponds to the $M_\pi = 220\,\MeV$
    ($310\,\MeV$).}
  \label{fig:h_A1-1}
\end{figure}

In Fig.~\ref{fig:h_A1-1}, the zero-recoil form factor $h_{A_1}(1)$ is
compared for different orders of current improvement.
All of the results are obtained by assuming that the matching factor
$\rho_{A_1}=1$ (\textit{i.e.} $\rho_{A_1}$ is blinded).
The size of the leading order $\mathcal{O}(\lambda^1)$ correction to
$h_{A_1}(1)$ is small, as is expected from HQET.
However, the second order correction of $\mathcal{O}(\lambda^2)$ is
larger than the leading order correction. This is unexpected from the
HQET.
Thus, the current improvement up to $\mathcal{O}(\lambda^2)$ seems
crucial, and a higher order improvement is interesting to check the
convergence of HQET expansion. The third order $\mathcal{O}
(\lambda^3)$ improved current is being implemented.

Fig.~\ref{fig:h_A1-1} shows that the $h_{A_1}(1)$ decreases by less
than $1\sigma$ as the pion mass changes from $310\,\MeV$ to
$220\,\MeV$ and by about $0.5\sigma$ as the lattice spacing changes
from $a=0.12\,\fm$ to $a=0.09\,\fm$.
Since these differences are not significant from the standpoint of
statistics, we need more data at various lattice spacings
(\textit{e.g.} superfine and ultrafine ensembles of the MILC HISQ
ensembles) in order to address the issue properly.
In addition, the change pattern with respect to the lattice
spacing and pion mass are similar for all orders of the current
improvement through $\mathcal{O}(\lambda^n),\,(n=0,1,2)$.
Thus, it could be the light quark discretization that dominates the
dependence on the lattice spacing and pion masses.

In Fig.~\ref{fig:h_A1-1}, the blue cross symbol represents the result
extrapolated to the physical limit from FNAL/MILC calculation
\cite{Bailey:2014tva} that was done by using the Fermilab action for
charm and bottom valence quarks on $2+1\text{-flavor}$ MILC asqtad
staggered ensembles.
In Fig.~\ref{fig:h_A1-1}, the red cross symbol represents the
extrapolated result from HPQCD \cite{Harrison:2017fmw}.
The HPQCD calculation uses HISQ action for the light and charm valence
quarks, and NRQCD action for bottom valence quark on $2+1+1
\text{-flavor}$ MILC HISQ ensembles.

The matching factor is $\rho_{A_1}=1 + \mathcal{O}(\alpha_s)$.
The correction term is not yet included, but is expected to be a
subpercent effect because we anticipate a large cancellation among
current renormalization factors between $Z_{A}$'s in the numerator and
$Z_{V}$'s in the denominator in the double ratio.
%

\section{$B\to D\ell\nu$ Form Factors: $h_{\pm}(w)$}
\label{sec:h_pm}
The hadronic matrix element of the $\BtoD$ decay amplitude can be
expressed in terms of semileptonic form factors $h_{\pm} (w)$ as
follows,
\begin{align}
  \frac{\matrixe{D(M_D,\bm{p}^\prime)}{V_\mu}{B(M_B,\bm{0})}}
       {\sqrt{2M_D}\sqrt{2M_B}}
       &= \frac{1}{2}\left\{h_{+}(w) (v+v^\prime)_\mu
       + h_{-}(w)(v-v^\prime)_\mu\right\} \,,
\end{align}
where the four velocities $v=p/M_B=(1,\bm{0})$, $v^\prime =
p^\prime/M_D = (E_D/M_D,\bm{p}^\prime/M_D)$ and recoil parameter $w =
v\cdot v^\prime$.
\begin{figure}[t!]
  \begin{subfigure}{0.500\linewidth}
    \includegraphics[width=\linewidth]{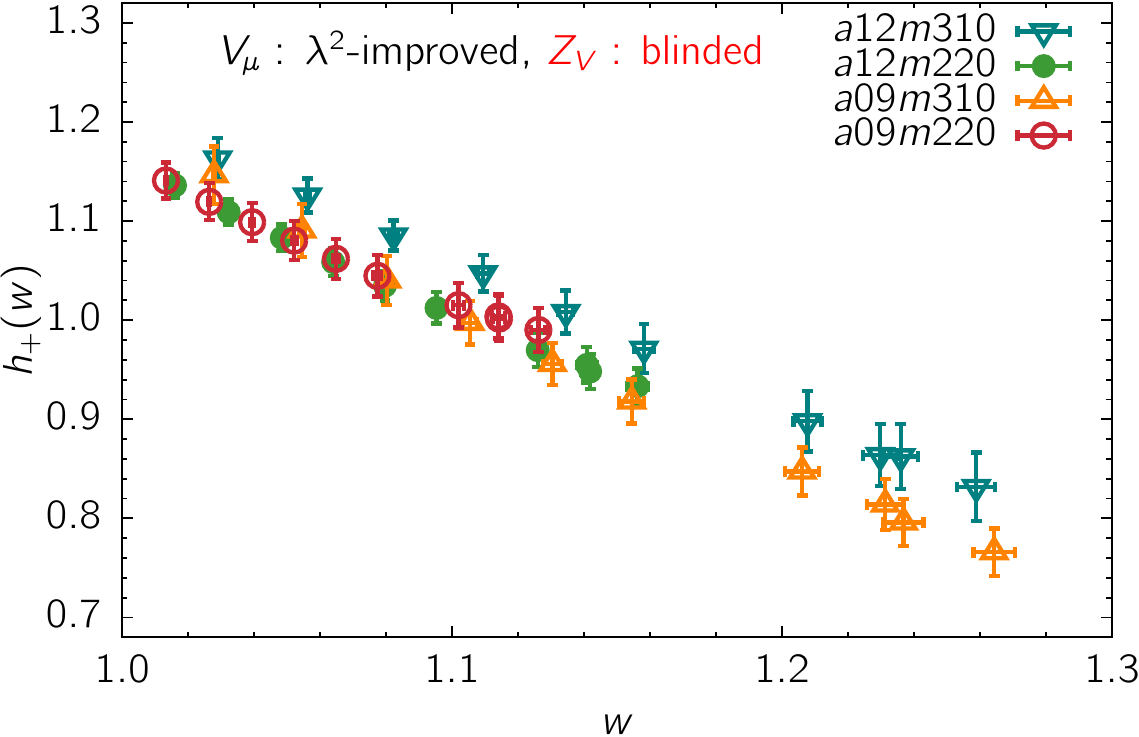}
    \caption{$h_+(w)$ vs. $w$}
    \label{fig:h_+:n}
  \end{subfigure}
  \hfill
  \begin{subfigure}{0.500\linewidth}   
    \includegraphics[width=\linewidth]{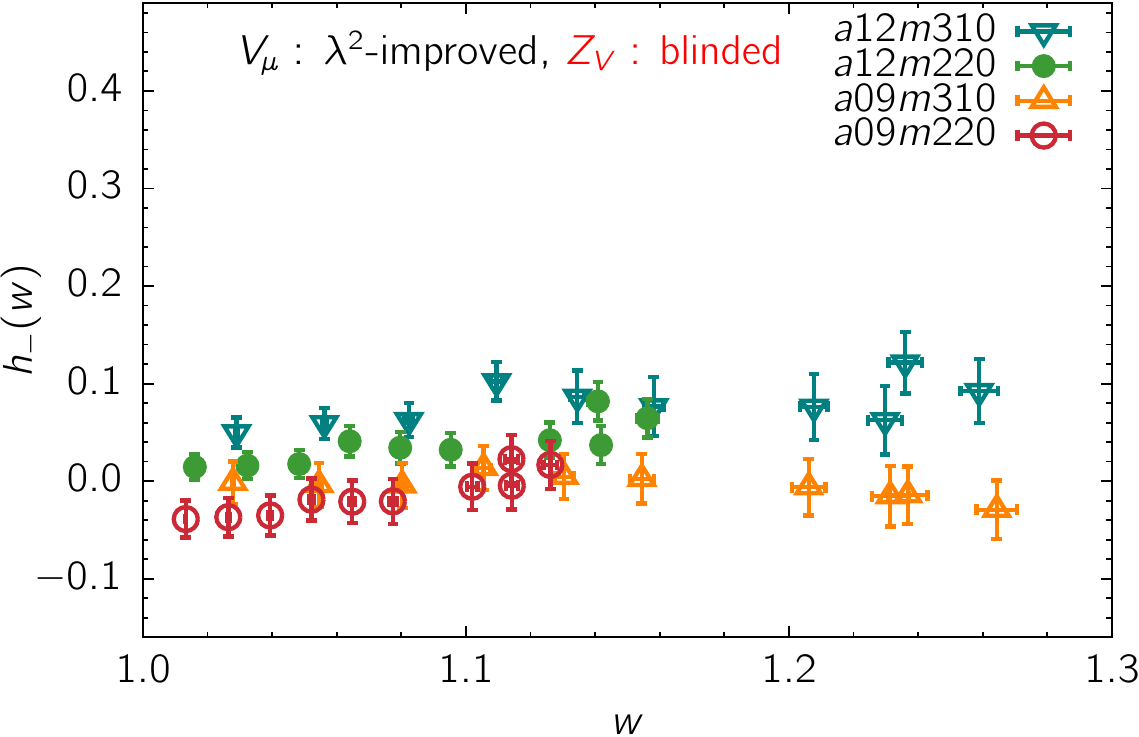}
    \caption{$h_-(w)$ vs. $w$}
    \label{fig:h_-:n}
  \end{subfigure}
  \caption{$h_{\pm}(w)$ as a function of recoil parameter $w$. The
    current is improved up to the $\lambda^2$ order.
  }
  \label{fig:h_pm:all}
\end{figure}
In Fig.~\ref{fig:h_pm:all}, we show form factors $h_\pm(w)$ as a
function of recoil parameter $w$ for the two coarse ($a12m310$ and
$a12m220$) and two fine ensembles ($a09m310$ and $a09m220$) of the
MILC HISQ lattices.
Here, we use the vector current improved up to the $\lambda^2$ order. 
Tiny variation with respect to light quark mass and lattice spacing
is observed for $h_+$ except the $a12m310$.
In contrast, the discretization effect is more visible for $h_-$.
\begin{figure}[b!]
  \begin{subfigure}{0.500\linewidth}
    \includegraphics[width=\linewidth]{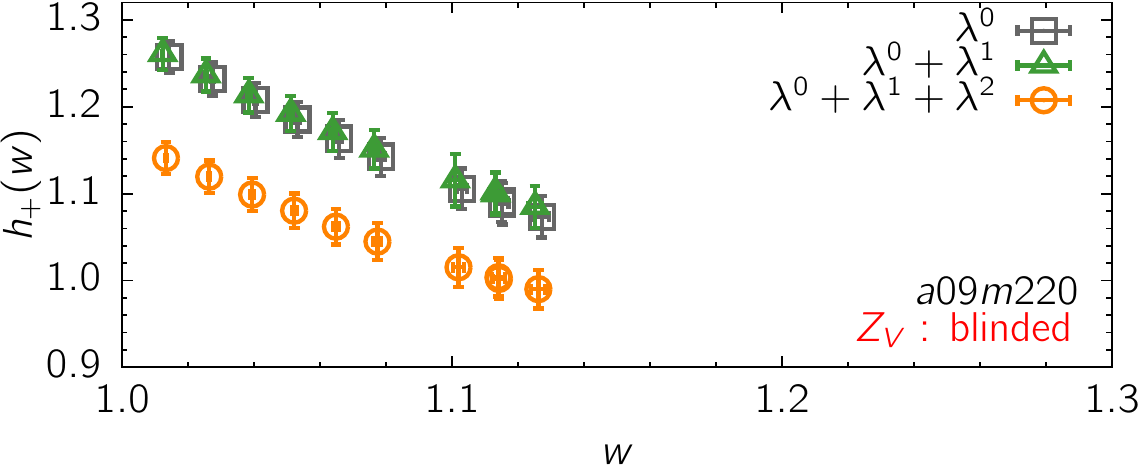}
    \caption{$h_+(w)$ vs. $w$}
    \label{fig:h_+:n}
  \end{subfigure}
  \hfill
  \begin{subfigure}{0.500\linewidth}
    \includegraphics[width=\linewidth]{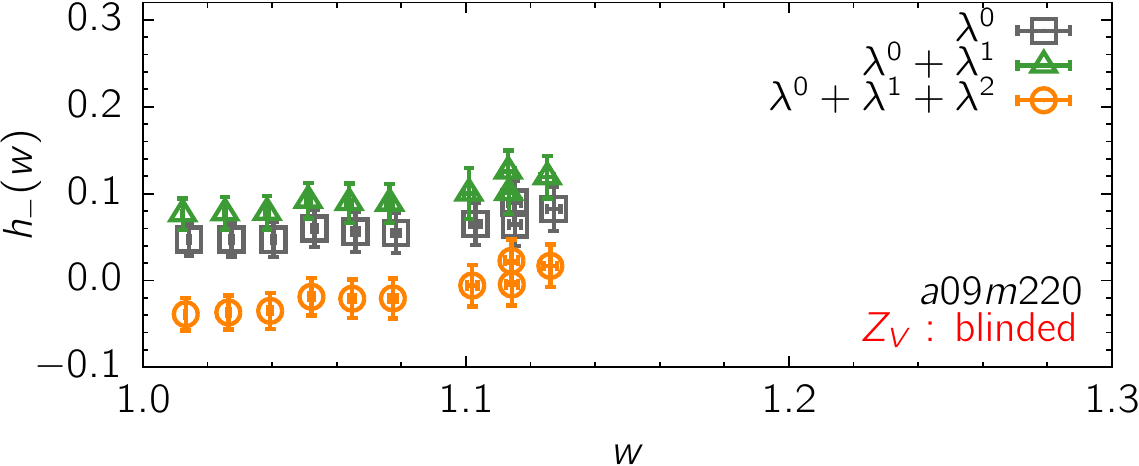}
    \caption{$h_-(w)$ vs. $w$}
    \label{fig:h_-:n}
  \end{subfigure}
  \caption{ Form factors $h_\pm(w)$ on $a09m220$ as a function of
    recoil parameter $w$. Here, we show how $h_\pm(w)$ changes as we
    improve the current up to the $\lambda^n$ order for $n=0,1,2$. }
  \label{fig:h_pm:n}
\end{figure}
In Fig.~\ref{fig:h_pm:n}, we present form factors $h_\pm(w)$ as a
function of recoil parameter $w$ for different orders of current
improvement on $a09m220$.
The first order correction ($\mathcal{O}(\lambda^1)$ in HQET) to the
unimproved ($\mathcal{O}{(\lambda^0)}$) current is negligible for
$h_{+}$ and small for $h_{-}$.
The second order correction of $\mathcal{O}(\lambda^2)$ reduces
$h_{+}$ about $10\%$ over the entire kinematic range.
The shift in $h_{-}$ is similar to that in $h_{+}$.
The same pattern of current improvement is found for all four
ensembles analyzed.

Non-perturbative calculation of renormalization factor $Z_{V}$ for
heavy quark current for $b \to c$ transition is being developed.
Tree-level renormalization $Z^\text{tree}_{V,cb} = \exp[(m_{1b}a +
  m_{1c}a)/2]$ is applied in this work.
%


\section{Summary \& Outlook}
\label{sec:sum}
It is crucial to improve the currents up to $\mathcal{O}(\lambda^3)$,
the same level as the OK action in our work.
We have implemented the $\mathcal{O}(\lambda^2)$ improvement in this
work and the $\mathcal{O}(\lambda^3)$ current improvement is being
implemented.
We plan to calculate the matching factors $\rho_{A_j}$ and $Z_V$ in
two independent ways: one is one-loop perturbation and the other is
non-perturbative renormalization using the RI-MOM and RI-SMOM schemes.

We plan to analyze two more data sets measured, $a12m130$ and
$a06m310$~\cite{Bhattacharya:2020ahq}, and to extend the measurement
to include other physical pion mass and finer lattices.
Statistics will be increased with the truncated solver method with
bias correction.
We plan to undertake the data analysis for the $B \to D^\ast\ell\nu$
decays soon.
%

%
%

\acknowledgments
\label{sec:ack}
%
We thank the MILC collaboration for providing the HISQ ensembles
to us.
Computations for this work were carried out in part on (i) facilities
of the USQCD collaboration, which are funded by the Office of Science
of the U.S. Department of Energy, (ii) the Nurion supercomputer at
KISTI and (iii) the DAVID GPU clusters at Seoul National University.
The research of W. Lee is supported by the Mid-Career Research Program
(Grant No.~NRF-2019R1A2C2085685) of the NRF grant funded by the Korean
government (MOE).
This work was supported by Seoul National University Research Grant in
2019.
W.~Lee would like to acknowledge the support from the KISTI
supercomputing center through the strategic support program for the
supercomputing application research (No.~KSC-2016-C3-0072,
KSC-2017-G2-0009, KSC-2017-G2-0014, KSC-2018-G2-0004,
KSC-2018-CHA-0010, KSC-2018-CHA-0043).
T.~Bhattacharya and R.~Gupta were partly supported by the
U.S. Department of Energy, Office of Science, Office of High Energy
Physics under Contract No.~DE-AC52-06NA25396.
S.~Park, T.~Bhattacharya, R.~Gupta and Y.-C.~Jang were partly
supported by the LANL LDRD program.
Y.-C.~Jang is partly supported by U.S.~Department of Energy under
Contract No.~DE-SC0012704.

\bibliographystyle{JHEP}

\bibliography{ref}

\end{document}